\documentclass[showpacs,preprintnumbers,amsmath,amssymb]{revtex4}
\usepackage{amsmath,amssymb,graphics,epsfig,subfigure}
\usepackage{color}

\begin{document}
\renewcommand{\baselinestretch}{1.3}

\title{General thermodynamic geometry approach for rotating Kerr anti-de Sitter black holes}

\author{Shao-Wen Wei \footnote{weishw@lzu.edu.cn, corresponding author},
        Yu-Xiao Liu \footnote{liuyx@lzu.edu.cn}}

\affiliation{Lanzhou Center for Theoretical Physics, Key Laboratory of Theoretical Physics of Gansu Province, School of Physical Science and Technology, Lanzhou University, Lanzhou 730000, People's Republic of China,\\
 Institute of Theoretical Physics $\&$ Research Center of Gravitation,
Lanzhou University, Lanzhou 730000, People's Republic of China,\\
 Academy of Plateau Science and Sustainability, Qinghai Normal University, Xining 810016, P. R. China}

\begin{abstract}
Combining with the small-large black hole phase transition, the thermodynamic geometry has been well applied to study the microstructure for the charged AdS black hole. In this paper, we extend the geometric approach to the rotating Kerr-AdS black hole and aim to develop a general approach for the Kerr-AdS black hole. Treating the entropy and pressure as the fluctuation coordinates, we construct the Ruppeiner geometry for the Kerr-AdS black hole by making the use of the Christodoulou-Ruffini-like squared-mass formula, which is quite different from the charged case. Employing the empirical observation of the corresponding scalar curvature, we find that, for the near-extremal Kerr-AdS black hole, the repulsive interaction dominates among its microstructure. While for far-from-extremal Kerr-AdS black hole, the attractive interaction dominates. The critical phenomenon is also observed for the scalar curvature. These results uncover the characteristic microstructure of the Kerr-AdS black hole. Such general thermodynamic geometry approach is worth generalizing to other rotating AdS black holes, and more interesting microstructure is expected to be discovered.
\end{abstract}



\maketitle

\section{Introduction}

Extended phase space thermodynamics has been one of the most active fields in the study of black holes. In this phase space, the cosmological constant is treated as an independent thermodynamic variable, the pressure, with its conjugate quantity being the thermodynamic volume \cite{Kastor}. This strengthens the understanding for the black hole thermodynamics. In particular, the precise comparison between the black hole and Van der Waals (VdW) fluid was completed \cite{Kubiznak}. Subsequently, the phase transition between different black hole phases was discovered. Much more interesting phase diagrams and critical phenomena were observed \cite{Kubiznak,Altamirano,AltamiranoKubiznak,Altamirano3,Dolan,Liu,Wei2,Frassino,
Cai,XuZhao,Hennigar,Hennigar2,Tjoa2,Ruihong}.

It is generally believed that phase transition is the result of the coexistence and competition among system microcomponents. So, the phase transition can be viewed as a probe to the microstructure of the thermodynamic system. For a black hole system, we have the well defined macroscopic thermodynamic quantities, such as the temperature and entropy. Then we can employ the phase transition to test black hole microstructure via the fluctuation theory.

One of the powerful approaches testing the microstructure of a thermodynamic system is the thermodynamic geometry, especially the Ruppeiner geometry \cite{Ruppeiner}. Starting with the Boltzmann entropy formula, the Ruppeiner geometry is well constructed. General understanding of the corresponding line element is that it measures the distance between two neighbouring fluctuation states. A longer distance indicates a less probable fluctuation. After applying the geometry to varies fluid systems, empirical observation states that there exists a significant understanding on the scalar curvature of the geometry. The positive or negative scalar curvature is produced by the repulsive or attractive interaction. This particular property rises the interest on the applications of the geometry. More importantly, the scalar curvature goes to negative infinity at the critical point of the phase transition. And the critical exponents were also obtained. This result suggests that the scalar curvature is linked with the correlation length of the system \cite{Ruppeiner}. Therefore, many microscopic properties can be reflected via the study of the Ruppeiner geometry \cite{Ruppeinerbb,Ruppeinercc,Dyjack}.

More than twenty years ago, the Ruppeiner geometry was introduced into the study of black holes \cite{CaiCho}. Early study focused on uncovering the divergent behavior of the heat capacity and the scalar curvature. It suggests that the information of the phase transition encoded in the heat capacity might also be revealed from the scalar curvature \cite{Luo,Sahaya,Ruppeinergg,MirzaZamani,Hosseini,Fazel,Zhangav}.

Although the scalar curvature of the Ruppeiner geometry has an important application to the phase transition, the first-order small-large black hole phase transition gains a well study until near ten years ago \cite{Kubiznak}. Actually, such phase transition was observed in \cite{Chamblin,Chamblin2,Cognola}. Thus the study of the small-large black hole phase transition gives us an opportunity to test the black hole microstructure in a geometric approach.

Combining with the phase transition, we first constructed a meaningful and well behaved Ruppeiner geometry by taking the pressure and mass as the fluctuation coordinates for the four-dimensional charged AdS black hole \cite{Weiw}. Novel microscopic property was uncovered. Further, we developed a more general geometric approach on testing black hole microstructure \cite{Weiwa2}. Excluding the coexistence region, where the equation of state might not applicable, we observed that the scalar curvature is always negative for the VdW fluid. This result indicates that there is only the attractive interaction between two fluid molecules, which is consistent with the rigid sphere model of the VdW fluid. However, when we applied the study to the charged AdS black hole, the result behaves quite differently. Beside the attractive interaction for the large black hole, we observed that the repulsive interaction also exists for the small black hole with higher temperature. This uncovers the general feature of the microstructure for the charged AdS black hole. Moreover, the critical behavior of the scalar curvature was also analytically studied. We confirmed that these results not only hold for the four-dimensional spacetime, but also for higher-dimensional cases \cite{Weiwa2,WeiWeiWei}.

The approach has also been generalized to other black hole backgrounds \cite{Dehyadegari,Zangeneh:2016fhy,Moumni,Miao,Du,Xuz,GhoshBhamidipati,Kumara2020,Yerra:2020oph,Wu:2020fij,
Vaid,Rizwan,Mansoori,Kumara,Mannw,Dehyadegariw,Wei2020d,YPhu,Xuzzm}. For the neutral AdS black hole, we constructed its Ruppeiner geometry, and found that there is only the attractive interaction for arbitrary spacetime dimension. However, when the black hole charge is present, the repulsive interaction will appear for the small black hole with higher temperature. So the charge is a key for the repulsive interaction. This conclusion not only holds in Einstein gravity, but also in modified gravities \cite{Yerra:2020oph,Wu:2020fij,Rizwan,YPhu}.

One of the interesting results was observed for the Gauss-Bonnet (GB) gravity \cite{Wei2020a}. For the five-dimensional neutral GB-AdS black hole, the attractive interaction dominates as expected. However, along the coexistence small and large black hole curves, we found that the scalar curvature of the Ruppeiner geometry keeps unchanged before and after the small-large black hole phase transition. This result suggests that the microscopic interaction can hold constantly even when the microstructure changes. Such particular property also holds for the charged black hole in the grand canonical ensemble \cite{Zhourun}. The similar result can also be found in Ref. \cite{Sahay} for the normal Hydrogen. However, in the canonical ensemble, this property violates, and the repulsive interaction emerges \cite{WeiLL}. The interesting dynamic process of the black hole phase transition in GB gravity can also be found in Refs. \cite{WeiLLa,WeiLLb}.

Combining with the phase transition, the geometric approach has been extensively applied to the charged AdS black hole case. However, for the rotating AdS black hole, no reference can be found. This is because that, different from the charged AdS black hole, the pressure cannot be expressed in terms of the temperature and volume for the rotating one, so it is impossible to obtain the critical point following the treatment of the charged AdS black hole. In order to further study the Ruppeiner geometry, one can expand the equation of state at the small angular momentum limit, and then obtain the critical point \cite{Gunasekaran}. However, one could not examine the critical phenomena by the approximate critical point. On the other hand, we found that the critical point and the coexistence curve can be analytically obtained through the isobaric lines in the temperature and entropy diagram. This study was also extended to higher-dimensional singly spinning and double equal spinning Kerr-AdS black holes \cite{Wei6,Wei7}, as well as the Kerr-Newmann black hole \cite{Cheng2}. So, in this paper, we would like to focus on the Kerr-AdS black hole and construct a general geometric approach for the rotating AdS black hole.

In this paper, basing on the thermodynamics and phase transition of the Kerr-AdS black hole, we shall develop a general geometric approach for the rotating AdS black hole starting with the first law of a rotating black hole. The paper is organized as follows. In Sec. \ref{null}, we briefly review the phase transition of the Kerr-AdS black hole. Then, the phase diagrams and critical exponents are studied through adopting the entropy and pressure as the thermodynamic variables. In Sec. \ref{rgam}, we investigate the Ruppeiner geometry for the rotating Kerr-AdS black hole. We start with the first law and obtain a general line element of the Ruppeiner geometry for the rotating black hole. If the Christodoulou-Ruffini-like squared-mass formula is obtained, the explicit geometry will be determined. Basing on it, we calculate the scalar curvature for the Kerr-AdS black hole. The corresponding behavior is exhibited in detailed. The critical exponent is also numerically studied. At last, we summarize and discuss our results in Sec. \ref{Conclusion}.

\section{Thermodynamics and phase transition for Kerr-AdS black hole}
\label{null}

There are lots of papers concerning the phase transitions for the charged AdS black hole. However, only a few ones can be found for the rotating AdS black hole, i.e., the Kerr-AdS black hole. In order to complete our study, in this section, we would like to review the study of the thermodynamics and phase transition for the Kerr-AdS black hole following Ref. \cite{Wei6}. This will attribute to the general construction of the thermodynamic geometry. Further, the microstructure will be tested as expected.

\subsection{Thermodynamics}

The line element of the Kerr-AdS black hole reads
\begin{eqnarray}
 ds^{2}=-\frac{\Delta}{\rho^{2}}\bigg(dt-\frac{a\sin^{2}\theta}{\Xi}d\varphi\bigg)^{2}
        +\frac{\rho^{2}}{\Delta}dr^{2}+\frac{\rho^{2}}{1-a^{2}/l^{2}\cos^{2}\theta}d\theta^{2}\nonumber\\
        +\frac{(1-a^{2}/l^{2}\cos^{2}\theta)\sin^{2}\theta}{\rho^{2}}\bigg(adt-\frac{r^{2}+a^{2}}{\Xi}d\varphi\bigg)^{2},
\end{eqnarray}
where the metric functions are given by
\begin{eqnarray}
 \rho^{2}&=&r^{2}+a^{2}\cos^{2}\theta,\quad
 \Xi=1-\frac{a^{2}}{l^{2}},\\
 \Delta&=&(r^{2}+a^{2})(1+r^{2}/l^{2})-2mr.
\end{eqnarray}
The temperature, entropy, and angular velocity read
\begin{eqnarray}
 T&=&\frac{r_{h}}{4\pi(r_{h}^{2}+a^{2})}
   \left(1+\frac{a^{2}}{l^{2}}+3\frac{r_{h}^{2}}{l^{2}}-\frac{a^{2}}{r_{h}^{2}}\right),\\
 S&=&\frac{\pi (r_{h}^{2}+a^{2})}{\Xi},\quad \Omega=\frac{a\Xi}{r_{h}^{2}+a^{2}}+\frac{a}{l^{2}}.
\end{eqnarray}
The horizon radius $r_{h}$ of the black hole can be obtained by solving $\Delta(r_{h})$=0. The black hole mass $M$ and angular momentum $J$ are related to the parameters $m$ and $a$ as
\begin{eqnarray}
 M=\frac{m}{\Xi^{2}},\quad J=\frac{am}{\Xi^{2}}.
\end{eqnarray}
In the extended phase space, the black hole mass $M$ is treated as the enthalpy $H$ rather than the internal energy \cite{Kastor}. The Gibbs free energy $G=H-TS$ of the black hole is
\begin{eqnarray}
 G=\frac{a^4 \left(r_{h}^2-l^{2}\right)+a^2\left(3l^{4}
   +2l^{2}r_{h}^2+3r_{h}^4\right)+l^{2} r_{h}^2
   \left(l^{2}-r_{h}^2\right)}{4 r_{h}\left(a^2-l^{2}\right)^2}.
\end{eqnarray}
In Ref. \cite{Wei6}, we have expressed the temperature, Gibbs free energy, and thermodynamic volume in terms of $S$, $J$, and $P$ as
\begin{eqnarray}
 T&=&\frac{S^2 \left(64 P^2S^2+32PS+3\right)
     -12\pi^2 J^2}{4\sqrt{\pi} S^{3/2} \sqrt{8 P S+3} \sqrt{12 \pi^2 J^2+S^2(8 P S+3)}},\label{TT}\\
 G&=&\frac{12\pi^2 J^2(16PS+9)-64 P^2 S^4+9S^2}{12\sqrt{\pi}\sqrt{S}\sqrt{8PS+3}
    \sqrt{12\pi^2J^2+S^2 (8PS+3)}},\\
 V&=&\frac{4\sqrt{S}\left(6\pi^2J^2+S^2(8PS+3)\right)}{3\sqrt{\pi}\sqrt{8PS+3}
   \sqrt{12\pi^2J^2+S^2(8PS+3)}}.\label{VV}
\end{eqnarray}
The specific volume can be determined with $v=2(3V/4\pi)^{1/3}$. The AdS radius $l$ has been replaced with the pressure via $P=3/8\pi l^2$. Employing these thermodynamic quantities, one can check that the following first law and Smarr law hold
\begin{eqnarray}
 dH&=&TdS+\Omega dJ+VdP,\label{firlaw}\\
 H&=&2TS+2\Omega J-2PV.
\end{eqnarray}
Moreover, we would like to show another interesting relation between these thermodynamic quantities, the Christodoulou-Ruffini-like squared-mass formula \cite{Cognola},
\begin{eqnarray}
 H^2=\frac{S}{4\pi}+\frac{\pi J^2}{S}+\frac{8\pi J^2P}{3}+\frac{4PS^2}{3\pi}+\frac{16P^2S^3}{9\pi}.\label{crlsm}
\end{eqnarray}
This relation is ignored in previous study, but we will show that this relation is very useful on constructing the Ruppeiner geometry for the Kerr-AdS black hole.

\subsection{Critical point and phase diagram}

In order to examine the VdW-like phase transition, one usually needs to study the behavior of the isothermal curve in the $P$-$V$ plane. However, from Eqs.~(\ref{TT}) and (\ref{VV}), we find that it is impossible to express the pressure $P$ in terms of $T$, $V$, and $J$. It should be noted that, following \cite{Gunasekaran}, one can obtain an approximate equation of state in the limit of small angular momentum $J$. Then the critical point can be approximately derived.

In Ref. \cite{Wei6}, we reconsidered the concept of the critical point, and claimed that it actually a relation between two types of thermodynamic quantities. The first type is the universal parameters, such as the pressure, volume, and temperature, which are universal quantities for an ordinary thermodynamic system. Another type is called the characteristic parameters such as $a$ and $b$ of VdW fluid, which describe the characteristic properties of a thermodynamic system. Under this interpretation, the Kerr-AdS black hole is a single characteristic parameter system. At the critical point, all these values of the pressure, volume, and temperature only depend on the angular momentum. Combining with the dimensional analysis, the explicit forms of these universal quantities can be uniquely determined. Then the coefficients can be easily fixed by the numerical calculation via solving $(\partial_{V}P)_{T,J}=(\partial_{V,V}P)_{T,J}=0$. Finally an exact critical point can be obtained.

Through the first law of black hole thermodynamics, we found that the critical point can also be obtained by conditions $(\partial_{S}T)_{P,J}=(\partial_{S,S}T)_{P,J}=0$. Solving them, we obtained the analytical critical point for the rotating Kerr-AdS black hole by making use of (\ref{TT}) for the first time. Since the analytic form is complex, here we only give its approximate expression \cite{Wei6}
\begin{eqnarray}
 P_{c}=\frac{0.002857}{J},\quad
 T_{c}=\frac{0.041749}{\sqrt{J}},\quad
 V_{c}=115.796503J^{\frac{1}{3}}.
\end{eqnarray}
For the pressure and temperature are lower than their critical values, the isobaric curves in the $T$-$S$ plane exhibits a nonmonotonic behavior, and the Gibbs free energy shows a characteristic swallow tail behavior. Both of them indicate the coexistence of the first-order small-large black hole phase transition. Moreover, this phase transition is found to be the VdW type. The coexistence curve starts at the origin, then extends to higher pressure and temperature, and finally ends at a critical point, where the phase transition becomes a second-order one. In the reduced parameter space, the fitting form of the coexistence curve is given by \cite{Wei6}
\begin{eqnarray}
 \tilde{P}&=&0.718781
   \tilde{T}^2+0.188586 \tilde{T}^3+0.061488 \tilde{T}^4+0.022704
   \tilde{T}^5\nonumber\\
   &+&0.002340 \tilde{T}^6+0.010547 \tilde{T}^7-0.008649 \tilde{T}^8+0.005919
   \tilde{T}^9-0.001717 \tilde{T}^{10},  \label{coex4}
\end{eqnarray}
where a reduced thermodynamic quantity $\tilde{A}$ is defined as $\tilde{A}=A/A_{c}$ with $A_{c}$ being its critical value. Obviously, in the reduced parameter space, the coexistence curve is independent of the angular momentum. Moreover, the reduced equation of state will also be independent of the angular momentum. According to this special property, the reduced temperature $\tilde{T}$ depends on $\tilde{P}$ and $\tilde{S}$ only. Therefore, here we adopt the $\tilde{P}$-$\tilde{S}$ plane to describe the black hole phase structure. Although the phase diagrams has been well studied in $P$-$T$ and $T$-$V$ plane, we shall see there are some novel properties of the phase structure in the $\tilde{P}$-$\tilde{S}$.

\begin{figure}
\center{
\includegraphics[width=7cm]{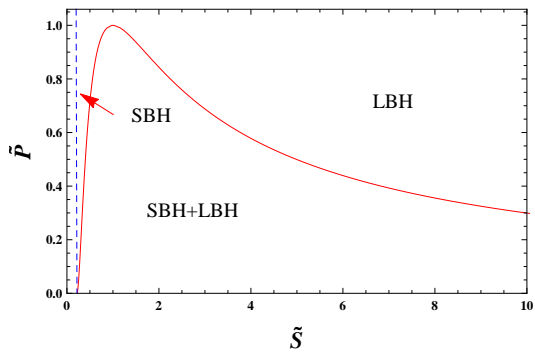}}
\caption{The phase diagram for the Kerr-AdS black holes shown in $\tilde{P}$-$\tilde{S}$ plane. ``SBH" and ``LBH" denote the small black hole and large black hole, respectively. The blue dashed curve denotes the extremal black hole with vanishing temperature, where the temperature vanishes. At the left side of the blue dashed curve, there is no black hole.}\label{pCP}
\end{figure}

Employing (\ref{coex4}), we exhibit the phase diagram for the Kerr-AdS black hole in Fig. \ref{pCP}. The red solid curve denotes the coexistence curve of the small and large black holes. Below the curve, it is the coexistence region. The small and large black holes are located at the left and right sides, respectively. For a fixed pressure $\tilde{P}$, we observe that the values of the reduced entropies for coexistence small and large black holes are different. This implies that the microstructure of the black hole undergoes a sudden change during the phase transition. In general, the microscopic interaction changes among the phase transition. However, one counter example is the five-dimensional GB black hole \cite{Wei2020a}, its interaction keeps the same even the microstructure of the black hole changes during the phase transition.

In addition, in the $\tilde{P}$-$\tilde{S}$ plane, there exists a region, where the temperature is negative. Here, we solve the extremal black hole case with $\tilde{T}=0$, which gives
\begin{eqnarray}
 \tilde{P}=\frac{\sqrt{2.3213 \tilde{S}^2+0.3333}-3.0472 \tilde{S}}{\tilde{S}^2}.
\end{eqnarray}
We describe it in Fig. \ref{pCP} by the blue dashed curve. At the left side of it, the solution has a negative temperature, and thus it is the non-black hole case. We will exclude such solution in the following study.

\subsection{Critical phenomenon}

In the above subsection, we have shown that, with the increase of the temperature, the small-large black hole phase transition tends to and ends at a critical point, where it is a second-order one. Here we would like to examine the critical phenomenon near that critical point.

\begin{figure}
\center{\subfigure[]{\label{Hpt}
\includegraphics[width=7cm]{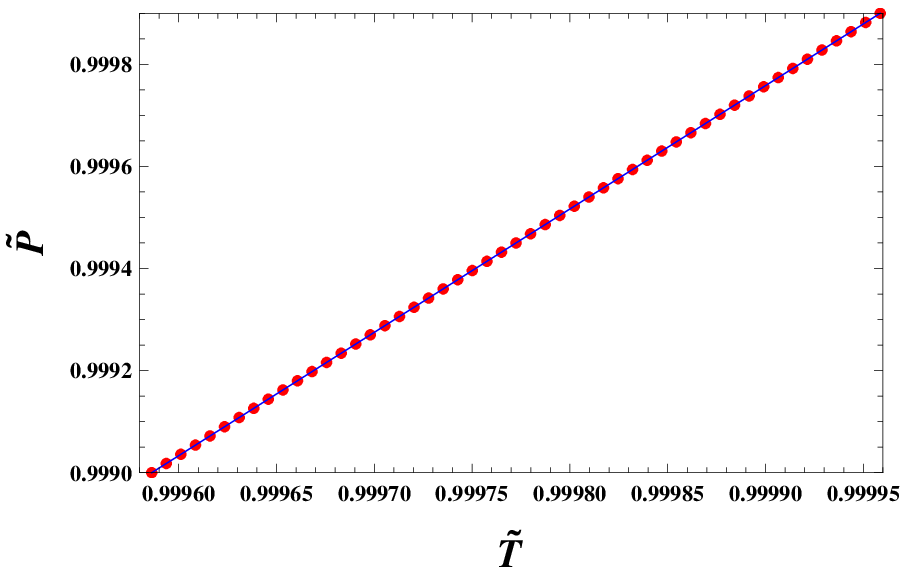}}
\subfigure[]{\label{HDS}
\includegraphics[width=7cm]{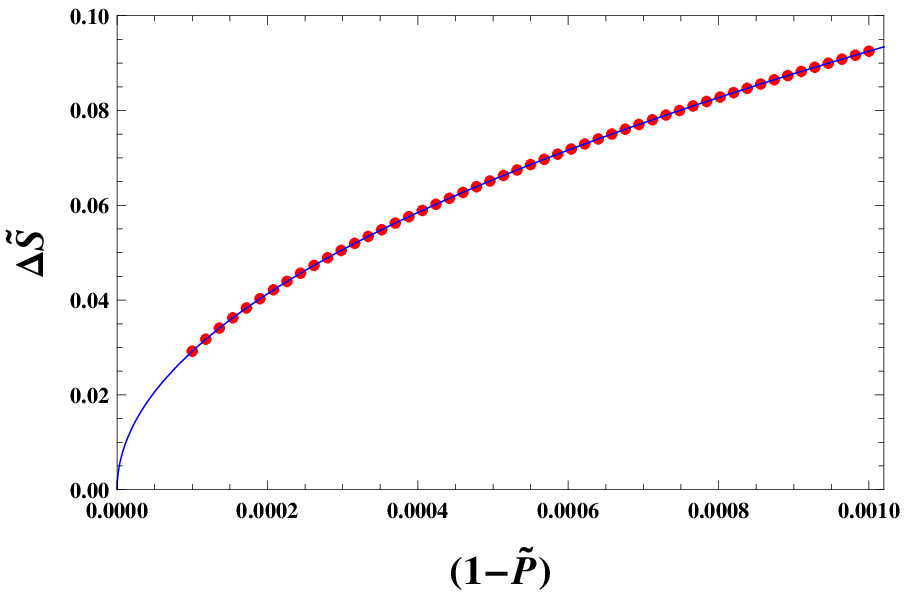}}}
\caption{The behaviors of the thermodynamic quantities near the critical point. The red dots denote the numerical results, and the blue curves are for the fitting results. (a) $\tilde{P}$ vs. $\tilde{T}$. (b) $\Delta\tilde{S}$ vs. ($1-\tilde{P}$).}\label{pHDS}
\end{figure}

Near the critical point, we show the numerical results (described by the red dots) in Fig. \ref{pHDS}. In Fig. \ref{Hpt}, the pressure $\tilde{P}$ is plotted as a function of $\tilde{T}$ for the phase transition. The result suggests $\tilde{P}$ is a linear function of $\tilde{T}$. Moreover, we also exhibit the deference $\Delta S$ between the coexistence small and large black holes in Fig. \ref{HDS}. Here we fit these data with the following forms
\begin{eqnarray}
 \tilde{P}&=&\alpha \tilde{T}+\beta,\\
 \Delta \tilde{S}&=&\gamma (1-\tilde{P})^{\delta},
\end{eqnarray}
where the fitting coefficients are given by
\begin{eqnarray}
 \alpha=2.4166,\quad \beta=-1.4166,\quad \gamma=2.9358,\quad \delta=0.5005.
\end{eqnarray}
In Fig. \ref{pHDS}, we also plot the fitting relation by the blue curves. It is easy to see that the numerical results and the fitting results are highly consistent with each other. Therefore, near the critical point, $\tilde{P}$ is linear with respect to $\tilde{T}$. Moreover, we have $\delta=0.5005$, which indicates that $\Delta \tilde{S}$ has a critical exponent $\frac{1}{2}$, and thus $\Delta \tilde{S}$ can be treated as an order parameter to characterize the small-large black hole phase transition.

In the previous study, one usually examines the critical exponents for the charged AdS black hole by employing the isothermal in the $P$-$V$ plane. However, this does not work for the Kerr-AdS black hole. Here we expect to develop a general study for obtaining the critical exponents. Before carrying out the analysis, we define the following parameters
\begin{eqnarray}
 t=1-\tilde{T}, \quad p=1-\tilde{P}, \quad s=1-\tilde{S}.
\end{eqnarray}
Below the critical point, $t$ and $p$ are positive, while $s$ is positive or negative for coexistence small or large black hole. Especially, near the critical point, they are all infinitesimal quantities. Expanding (\ref{TT}) near the critical point, we have
\begin{eqnarray}
 t=a_{03}s^{3}+(a_{10}+a_{11}s+a_{12}s^2)p+\mathcal{O}(ps^3,s^4).\label{expaing}
\end{eqnarray}
This expanding form is actually a general result for the phase transition. For the Kerr-AdS black hole, these coefficients are
\begin{eqnarray}
 a_{03}=0.097147,\quad a_{10}=0.413637,\quad
 a_{11}=-0.207402,\quad a_{12}=-0.052866.
\end{eqnarray}
In Ref. \cite{Wei3}, we showed that the equal area law holds in the $\tilde{T}$-$\tilde{S}$ plane for a fixed pressure $P$ and angular momentum $J$, which reads
\begin{eqnarray}
 \oint SdT=0,
\end{eqnarray}
with $T_*$ being the phase transition temperature. Near the critical point, $T$ is a single valued function of $S$, so we have
\begin{eqnarray}
 \int_{S_s(T_*)}^{S_l(T_*)} S\frac{dT}{dS}dS=0,
\end{eqnarray}
where $S_s(T_*)$ and $S_l(T_*)$ are the entropy for the coexistence small and large black holes, respectively. In terms of these infinitesimal quantities, we obtain
\begin{eqnarray}
 \int_{s_s}^{s_l} (1-s)\frac{dt}{ds}ds=0.
\end{eqnarray}
Combining with (\ref{expaing}) and carrying out the integral, we have
\begin{eqnarray}
 s_s=\sqrt{-\frac{a_{11}}{a_{03}}}*p^{\frac{1}{2}},\quad
 s_l=-\sqrt{-\frac{a_{11}}{a_{03}}}*p^{\frac{1}{2}}.
\end{eqnarray}
So, at the critical point, both $s_s$ and $s_l$ have universal critical exponent $\frac{1}{2}$. In addition,
\begin{eqnarray}
 \Delta s=s_l-s_s=-2\sqrt{-\frac{a_{11}}{a_{03}}}*p^{\frac{1}{2}}.
\end{eqnarray}
Clearly, $\Delta s$ also has the critical exponent $\frac{1}{2}$, which is consistent with the numerical result that $\delta=0.5005$. Further, by setting $s=0$ in (\ref{expaing}), we find
\begin{eqnarray}
 \left.\frac{dp}{dt}\right|_{s=0}=\frac{1}{a_{10}}.
\end{eqnarray}
Since $\frac{dp}{dt}=\frac{d\tilde{P}}{d\tilde{T}}$, one must have $\alpha=1/a_{10}$. When substituting the values of $\alpha$ and $a_{10}$ into it, it is easy to check that this relation holds. Employing the Clapeyron equations $\frac{dP}{dT}=\frac{\Delta S}{\Delta V}$, we can also find
\begin{eqnarray}
 \Delta \tilde{V}=-2\frac{S_{c}}{V_{c}}\sqrt{-\frac{a_{11}}{a_{10}^2a_{03}}}*p^{\frac{1}{2}}.
\end{eqnarray}
At the critical pressure $p$=0, it is easy to get the following relation
\begin{eqnarray}
 t\sim s^{3}.
\end{eqnarray}
Actually, other critical exponents can also be obtained via the expanding form (\ref{expaing}).

\section{Ruppeiner geometry and microstructure}
\label{rgam}

In this section, we would like to construct the Ruppeiner geometry and uncover the special property of the Kerr-AdS black hole black hole microstructure.

\subsection{First law and Ruppeiner geometry}

For the charged AdS black hole, we have constructed the generic Ruppeiner geometry, where the temperature $T$ and thermodynamic volume $V$ are the fluctuation coordinates. When one knows the equation of state $P=P(T,V)$, the Ruppeiner geometry will be given. Such approach has been extensively applied to different black holes, and interesting black hole microstructure is revealed. However, for the Kerr-AdS black hole, as shown above, the pressure cannot be expressed as an explicit function of $T$ and $V$. So we need to develop another approach.

Since the thermodynamic quantities can be expressed as $P$, $S$, and $J$, we adopt $P$ and $S$ as the fluctuation coordinates in our new approach. Now we start with the first law. Via (\ref{firlaw}), we obtain the first law
\begin{eqnarray}
 dU=d(H-PV)=TdS+\Omega dJ-PdV,\label{modflaw}
\end{eqnarray}
where $U$ is the internal energy of the black hole system. For a fixed angular momentum $J$, we obtain
\begin{eqnarray}
 dS=\frac{1}{T}dU+\frac{P}{T}dV.
\end{eqnarray}
Then the line element of the Ruppeiner geometry could be expressed as
\begin{eqnarray}
 ds^2=-d\left(\frac{1}{T}\right)dU-d\left(\frac{P}{T}\right)dV.
\end{eqnarray}
Inserting (\ref{modflaw}) into the above equation, one easily gets
\begin{eqnarray}
 ds^2=\left(\frac{1}{T}\right)dTdS-\left(\frac{1}{T}\right)dPdV.
\end{eqnarray}
By making the use of the following transformations
\begin{eqnarray}
 dT&=&\left(\frac{\partial T}{\partial S}\right)_PdS
 +\left(\frac{\partial T}{\partial P}\right)_SdP,\\
 dV&=&\left(\frac{\partial V}{\partial S}\right)_PdS
 +\left(\frac{\partial V}{\partial P}\right)_SdP,
\end{eqnarray}
we approach
\begin{eqnarray}
 ds^2=\frac{1}{T}\left(\frac{\partial T}{\partial S}\right)_PdS^2
      +\frac{1}{T}\left(\frac{\partial T}{\partial P}\right)_SdSdP
      -\frac{1}{T}\left(\frac{\partial V}{\partial S}\right)_PdPdS
      -\frac{1}{T}\left(\frac{\partial V}{\partial P}\right)_SdP^2.
\end{eqnarray}
From (\ref{firlaw}), we have
\begin{eqnarray}
 \left(\frac{\partial T}{\partial P}\right)_S=\left(\frac{\partial V}{\partial S}\right)_P.
\end{eqnarray}
So, the element of the Ruppeiner geometry can be simplified as
\begin{eqnarray}
 ds^2=\frac{1}{T}\left(\frac{\partial T}{\partial S}\right)_PdS^2
 -\frac{1}{T}\left(\frac{\partial V}{\partial P}\right)_SdP^2.
\end{eqnarray}
Using Eqs. (\ref{TT}) and (\ref{VV}), the metric will be obtained. Furthermore, via (\ref{firlaw}), the line element has another form
\begin{eqnarray}
 ds^2=\left(\frac{\partial H}{\partial S}\right)^{-1}\left(\frac{\partial^2 H}{\partial S^2}\right)_PdS^2
 -\left(\frac{\partial H}{\partial S}\right)^{-1}\left(\frac{\partial^2 H}{\partial P^2}\right)_SdP^2.\label{lineele}
\end{eqnarray}
So, for a Kerr-AdS black hole, if one knows the Christodoulou-Ruffini-like squared-mass formula (\ref{crlsm}), the Ruppeiner geometry will be given directly. This is also an important application of the squared-mass formula.

\subsection{Scalar curvature}

\begin{figure}
\subfigure[]{\label{RJ08a}
\includegraphics[width=7cm]{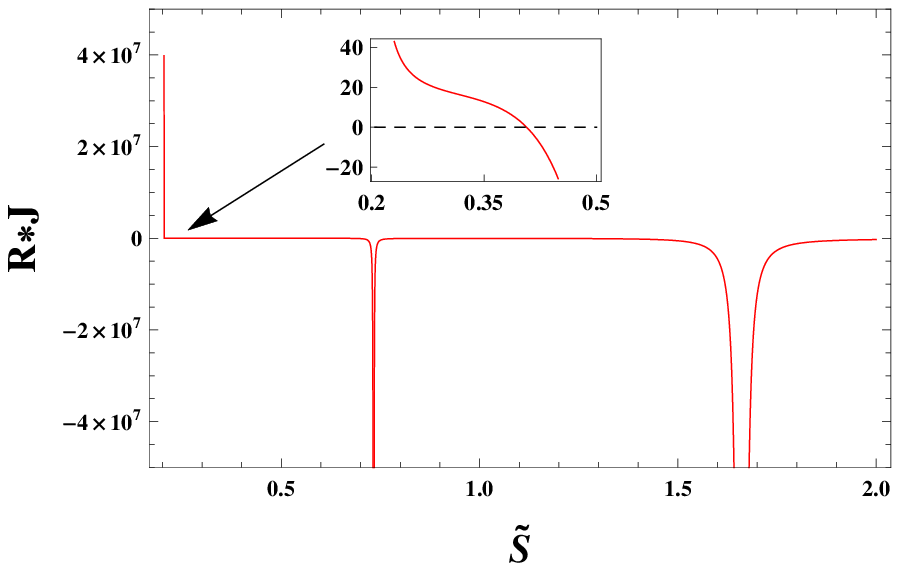}}
\subfigure[]{\label{RJ09b}
\includegraphics[width=7cm]{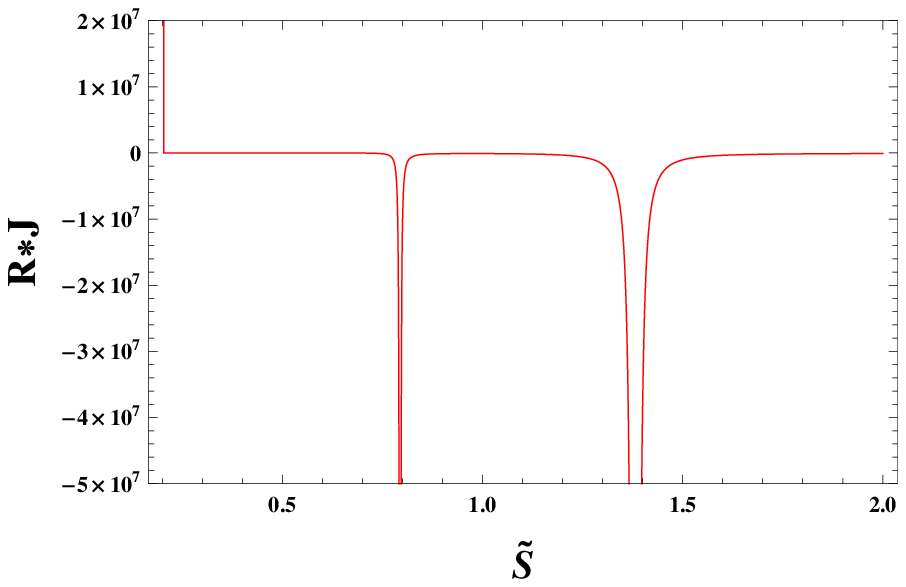}}
\subfigure[]{\label{RJ10c}
\includegraphics[width=7cm]{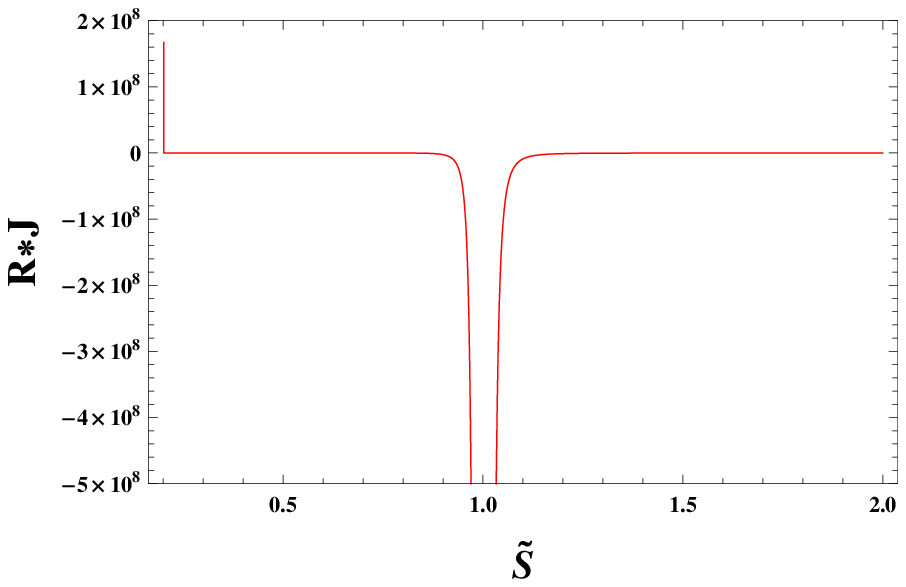}}
\subfigure[]{\label{RJ12d}
\includegraphics[width=7cm]{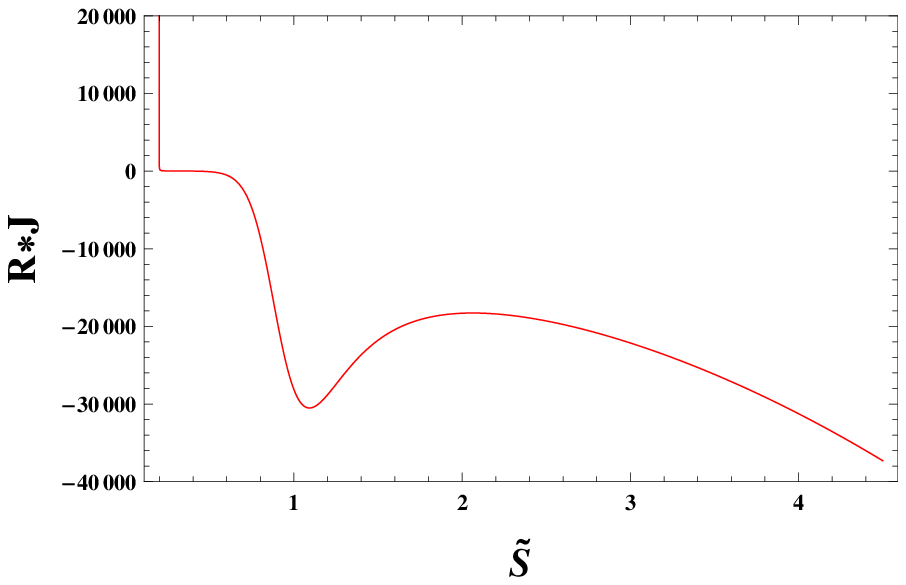}}
\caption{The behavior of the scalar curvature $R*J$ as a function of $\tilde{S}$ for fixed $\tilde{P}$. The solution with negative temperature has been excluded. (a) $\tilde{P}$=0.8. (b) $\tilde{P}$=0.9. (c) $\tilde{P}$=1.0. (d) $\tilde{P}$=1.2.}\label{pRJ12d}
\end{figure}

By using the Christodoulou-Ruffini-like squared-mass formula (\ref{crlsm}), we can calculate the scalar curvature through (\ref{lineele}). The non-vanishing components of the metric are given by
\begin{eqnarray}
 g_{SS}&=&-\frac{3 S (4 P S+1)}{12 \pi ^2 J^2+S^2 (8 P
   S+3)}+\frac{2 S (16 P S (8 P S+3)+3)}{S^2 (8 P
   S+1) (8 P S+3)-12 \pi ^2 J^2}-\frac{4 P}{8 P
   S+3}-\frac{3}{2 S},\\
 g_{PP}&=&\frac{1536 \pi ^4 J^4 S^3}{(8 P S+3) \left[12 \pi ^2
   J^2+S^2 (8 P S+3)\right] \left[S^2 (8 P S+1) (8 P
   S+3)-12 \pi ^2 J^2\right]}.
\end{eqnarray}
Following the construction of the Riemannian geometry, we obtain the scalar curvature $R$ of the Ruppeiner geometry for the Kerr-AdS black hole, which is given by
\begin{eqnarray}
 R=-\frac{2 A^2 C \left(A S^2+12 \pi ^2
   J^2\right)}{3 \pi ^4 B^2 J^4 S \left(12 \pi ^2
   J^2-A S^2 (8 P S+1)\right)},\label{scrg}
\end{eqnarray}
where
\begin{eqnarray}
 A&=&3 + 8 P S,\\
 B&=&S^4 (-1 + 8 P S) A^3 + 24 J^2 \pi^2 S^2 A^2 (3 + 16 P S) +
 144 J^4 \pi^4 (9 + 32 P S),\\
 C&=&-4 A^6 P S^{13}+3 \pi ^2 A^5 J^2 S^{10} [8 P S (24 P
   S+1)-3]-24 \pi ^4 A^3 J^4 S^8 \left[4 P S
   \left(1536 P^3 S^3-440 P
   S-59\right)+33\right]\nonumber\\
   &&+144 \pi ^6 A^2 J^6 S^6 [64
   P S (P S+2) (24 P S (8 P S+7)+47)+447]
   +41472 \pi^{10} A J^{10} S^2 (52 P S+17)\nonumber\\
   &&+1728 \pi ^8 A J^8
   S^4 [8 P S (16 P S (60 P S+67)+375)+333]+248832
   \pi ^{12} J^{12}.
\end{eqnarray}
We describe the scalar curvature $R$ in Fig. \ref{pRJ12d}. Taking Fig. \ref{RJ08a} as an example, where $\tilde{P}$=0.8, we observe one positive divergence and two negative divergences. The positive one is resulted by the vanishing temperature. So for the extremal black hole, $R=+\infty$. This phenomenon can also be found in other three figures. Besides, near the extremal black hole case, $R$ is positive, which indicates that there is the repulsive interaction among the micromolecules of the black hole. For the far-from-extremal black hole, $R$ is negative. Moreover, two negative divergent points are observed at certain $\tilde{S}$. With the increase of $\tilde{P}$, these two divergent points get close. At the critical case in Fig. \ref{RJ10c}, these two points merge, and only one divergent point is present. For the reduced pressure $\tilde{P}$ is beyond its critical value (see Fig. \ref{RJ12d}), the divergent behavior disappears and only a local well leaves near $\tilde{S}$=1.

\begin{figure}
\center{
\includegraphics[width=7cm]{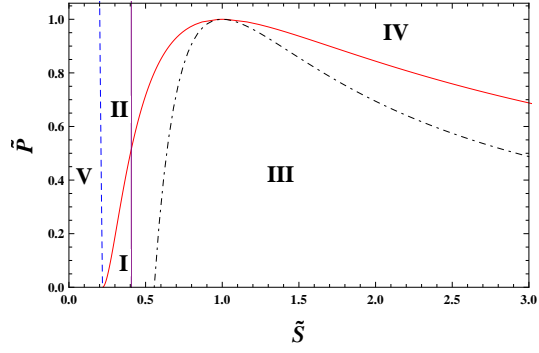}}
\caption{The sign ranges of $R$. The red solid curve is the coexistence curve. The purple line and the black dot dashed curve correspond to zero points and divergent points of $R$. The left blue dashed line denotes the extremal black hole case.}\label{PHS4}
\end{figure}

Furthermore, we obtain the zero points and divergent points of $R$ described respectively by the purple line and the black dot dashed curve in Fig. \ref{PHS4}. The red curve is for the coexistence curve. The phase space is divided into five regions. Region V is for the non-black hole case, where the solution has negative temperature, and it will not be considered here. $R$ is positive for regions I and II, while negative for other regions. Since in the coexistence region, the equation of state (\ref{TT}) may not hold, the scalar curvature (\ref{scrg}) does not applicable. We exclude region I and III. As a result, other calculation is only applicable in regions II and IV. Region II is near the extremal black hole line. Hence it denotes the region for the near-extremal black hole. In this region, $R$ is positive, which implies that there is a repulsive interaction among the microstructure. In region IV, the negative $R$ shows that there is an attractive interaction. Comparing with the charged AdS black hole, we find that the angular momentum $J$ has the same influence as the charge, i.e., the positive $R$ is present and repulsive interaction exists.

\begin{figure}
\center{\subfigure[]{\label{RJCSa}
\includegraphics[width=7cm]{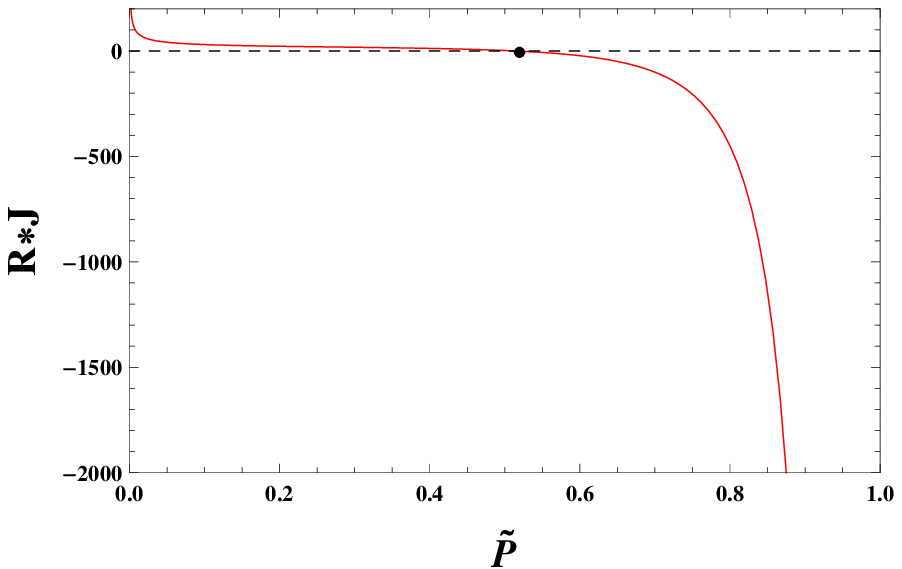}}
\subfigure[]{\label{RJCLb}
\includegraphics[width=7cm]{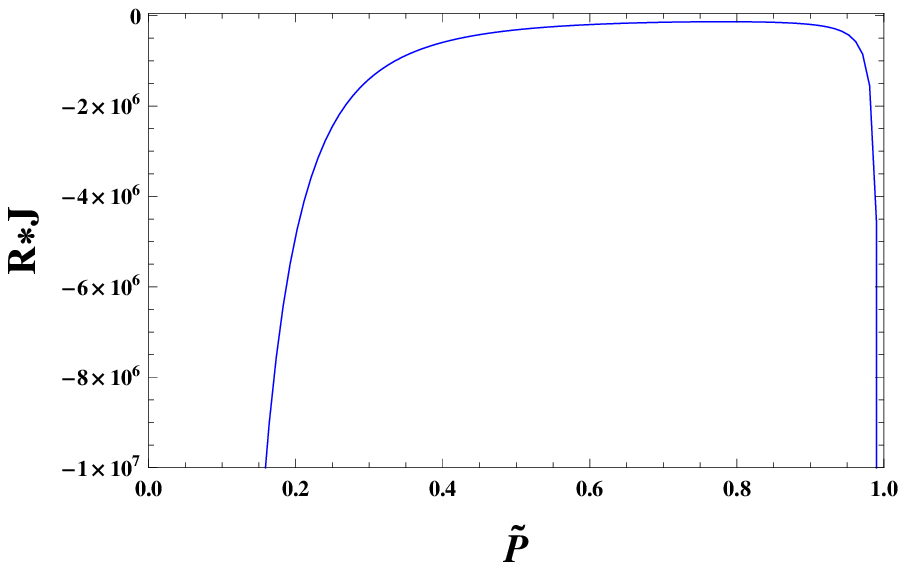}}}
\caption{The scalar curvature $R*J$ as a function of the reduced pressure $\tilde{P}$. Black dot denotes the zero point of $R$, which is located at $\tilde{P}$=0.5151. (a) Along the coexistence small black hole. (b) Along the coexistence large black hole.}\label{pRJCLb}
\end{figure}

The coexistence curve is an important curve in the black hole phase transition. We focus on the variation of $R$ along the curve. Varying $\tilde{P}$ from 0 to 1, we show the behavior $R*J$ in Fig. \ref{pRJCLb}. Along the coexistence small black hole curve, we observe that $R*J$ diverges both at $\tilde{P}$=0 and 1. At $\tilde{P}$=0, it has a positive divergent behavior, which is actually caused by the $T$=0. With the increase of $\tilde{P}$, it decreases and crosses the $\tilde{P}$-axis at $\tilde{P}$=0.5151. At last, it goes to negative infinity at the critical pressure. Such pattern suggests that the repulsive interaction exists for the low pressure black hole. For the coexistence large black hole, $R$ behaves quite differently. $R*J$ goes to negative infinity at both $\tilde{P}$=0 and 1. As we know, for the coexistence large black hole, low $\tilde{P}$ corresponds to the super-large black hole. For such black hole, $R$ has a large negative value indicating there is a strong attractive interaction. At the critical pressure, $R$ also negatively diverges as that of the coexistence small black hole.

\subsection{Critical phenomenon}

In the last subsection, we show that $R$ goes to negative infinity both for the coexistence small and large black holes. Here we attempt to numerically study such behavior.

\begin{figure}
\center{
\includegraphics[width=7cm]{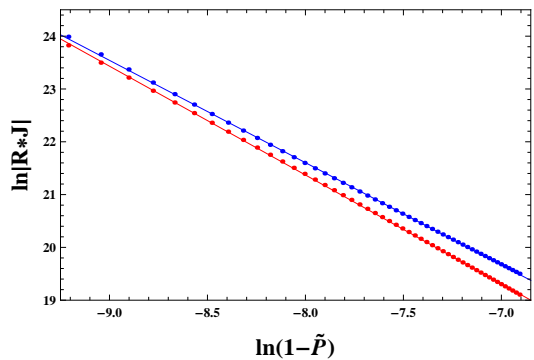}}
\caption{The numerical and fitting results of $R*J$ near the critical point marked with the dots and solid lines, respectively. The bottom red line and dots are for the coexistence small black hole, and the top blue line and dots are for the coexistence large black hole.}\label{RFit}
\end{figure}

We suppose that the scalar curvature has the following behavior at the critical point
\begin{eqnarray}
 R*J\propto (1-\tilde{P})^{-b},
\end{eqnarray}
or
\begin{eqnarray}
 \ln |R*J| =-b\ln (1-\tilde{P})+d.
\end{eqnarray}
Then we numerically calculate the coexistence small and large black holes quite near the critical point. The numerical results are exhibited with the dots in Fig. \ref{RFit}. After fitting these data, we obtain the fitting coefficients
\begin{eqnarray}
 b&=&2.06, \quad d=4.88 \quad \mathrm{(CSBH)},\\
 b&=&1.94, \quad d=6.10 \quad \mathrm{(CLBH)},
\end{eqnarray}
where "CSBH" and "CLBH" are for the coexistence small and large black holes, respectively. From the above fitting results, we observe that $b$ is near 2 for both the small and large black holes. Therefore, it implies that at the critical point, $R$ has a critical exponent 2, which is similar to the charged AdS black hole. So the critical exponent of $R$ also holds for both the charged and rotating AdS black holes.

\section{Conclusions and discussions}
\label{Conclusion}

In this paper, we developed a new general thermodynamic geometry approach for the rotating AdS black holes. Then we applied it to the Kerr-AdS black hole. Some characteristic behaviors of the black hole microstructure were also disclosed.

We firstly reviewed the thermodynamics and phase transition for the Kerr-AdS black hole, and showed that the approach studying the phase transition for the charged AdS black hole could not be generalized to the Kerr-AdS black hole case. Then we suggested that the phase transition can be studied through the isobaric curves in the $T$-$S$ diagram. By constructing the equal area law, the coexistence small and large black hole curves were obtained. For the Kerr-AdS black hole, we found that the thermodynamics and phase transition can be well studied in the $S$-$P$ parameter space. Then the phase diagram was exhibited. The small and large black hole regions were given, as well as their coexistence region. In particular, the non-black hole region was also shown, where the solution has negative temperature. After expanding the temperature in terms of the entropy and pressure, we analytically obtained the critical exponents for the phase transition of the Kerr-AdS for the first time.

Secondly, we successfully constructed the Ruppeiner geometry by taking the entropy $S$ and pressure $P$ as the fluctuation coordinates. For a rotating AdS black hole, we proposed that if the Christodoulou-Ruffini-like squared-mass formula is given, the explicit line element of the Ruppeiner geometry (\ref{lineele}) can be obtained easily. Since the squared-mass formula  extensively exists for the rotating AdS black hole, our construct can be easily extended to other black holes.

Based on this general approach, we obtained the Ruppeiner geometry for the Kerr-AdS black hole. The scalar curvature was calculated. Similar to the charged AdS black hole, there are two negative divergent points of the scalar curvature when the pressure is below its critical value. At the critical case, these two divergent points merger. And no negative divergent behavior can be found above the critical point. Moreover, we observed a positive divergent point for an arbitrary pressure at small entropy, which actually corresponds to the extremal black hole with vanishing temperature. Combining with the empirical observation of $R$, we obtained the result that for the near-extremal black hole, the repulsive interaction dominates among its microstructure, while the attractive interaction dominates for the far-from-extremal black hole.

At last, we numerically confirmed that, near the critical point, $R$ has a critical exponent 2, which is the same value as that for the charged AdS black hole.

From previous study, it has been shown that the present of the black hole charge leads to the repulsive interaction. In this paper, we also showed that the angular momentum has the similar effect. Since in this paper, we developed a general approach to construct the Ruppeiner geometry for the rotating AdS black hole, it is valuable to extend the study to other rotating black holes, and we believe more features of the microstructure will be observed.

\section*{Acknowledgements}
This work was supported by the National Natural Science Foundation of China (Grants No. 12075103, No. 11675064, No. 11875151, and No. 12047501), the 111 Project (Grant No. B20063), and the Fundamental Research Funds for the Central Universities (No. Lzujbky-2019-ct06).

\end{document}